\documentclass[aps,prl,twocolumn,superscriptaddress,showpacs,amsmath,amssymb]{revtex4-1}

\usepackage{graphicx}

\begin{document}

\title{Real-time characterization of the mechanical behaviour of an actively growing bacterial culture by rheology}

\author{R. Portela}
\affiliation{Centro de Recursos Microbiol\'ogicos, Faculdade de Ci\^encias e Tecnologia, Universidade Nova de Lisboa, 2829-516 Caparica, Portugal}
\affiliation{Laborat\'orio de Gen\'etica Molecular, ITQB, Universidade Nova de Lisboa, 2780 Oeiras, Portugal}

\author{P. L. Almeida}
\affiliation{ISEL,
Rua Conselheiro Em\'{\i}dio Navarro 1, P-1959-007 Lisboa, Portugal}
\affiliation{CENIMAT/I3N, Faculdade Ci\^encias e Tecnologia, Universidade Nova de Lisboa, 2829-516 Caparica, Portugal}

\author{P. Patr\'{\i}cio}
\affiliation{ISEL,
Rua Conselheiro Em\'{\i}dio Navarro 1, P-1959-007 Lisboa, Portugal}
\affiliation{Centro de F\'{\i}sica Te\'orica e Computacional,
Universidade de Lisboa,
Av. Prof. Gama Pinto 2, P-1649-003 Lisboa, Portugal}

\author{T. Cidade}
\affiliation{CENIMAT/I3N, Faculdade Ci\^encias e Tecnologia, Universidade Nova de Lisboa, 2829-516 Caparica, Portugal}
\affiliation{Departamento de Ci\^encia dos Materiais, Faculdade de Ci\^encias e Tecnologia, Universidade Nova de Lisboa, 2829-516 Caparica, Portugal}

\author{R. G. Sobral}
\email{These authors contributed equally to this work. E-mail: rgs@fct.unl.pt; cleal@adf.isel.pt}
\affiliation{Centro de Recursos Microbiol\'ogicos, Faculdade de Ci\^encias e Tecnologia, Universidade Nova de Lisboa, 2829-516 Caparica, Portugal}

\author{C. R. Leal}
\email{These authors contributed equally to this work. E-mail: rgs@fct.unl.pt; cleal@adf.isel.pt}
\affiliation{ISEL,
Rua Conselheiro Em\'{\i}dio Navarro 1, P-1959-007 Lisboa, Portugal}
\affiliation{CENIMAT/I3N, Faculdade Ci\^encias e Tecnologia, Universidade Nova de Lisboa, 2829-516 Caparica, Portugal}

\date{\today}

\begin{abstract}
The population growth of a \textit{Staphylococcus aureus} culture was followed by rheological measurements, under
 steady-state and dynamic shear flows. We observed a rich viscoelastic behaviour as a consequence of the bacteria activity. First, the viscosity increased $\sim 10\times$ due to cell multiplication and aggregation. This viscosity increase presented several drops and full recoveries, which are reproducible, allowing us to evoke the existence of a percolation phenomenon. Eventually, as the bacteria population reached a final stage of development, fulfilling the sample volume, the viscosity returned to its initial value, most probably caused by a change in the bacteria physiological activity, in particular the decrease of their adhesion properties. Finally, the viscous and the elastic moduli presented power law behaviours compatible with the ``soft glassy materials'' model, which exponents are dependent on the bacteria growth stage.
\end{abstract}

\maketitle


The study of the mechanical properties of living bacteria in a liquid rich medium, environment commonly used in laboratorial settings, opens a new perspective on the bacterial physiology and behaviour during population growth.
The application of mechanical methods is a challenging approach, since many variables such as nutrient availability, nutrient diffusivity, cell growth rate or biomass density, among others, are known to impact and alter the growth behaviour and spatial structure of the population \cite{Nadell2010}.

Different rheological techniques have been developed to study living systems, both eukaryotic \cite{Fabry2001,Wilhelm2008} and prokaryotic
\cite{Klapper2002,Rogers2008,Rupp2005}, aiming to describe their mechanical response to different physical and/or chemical conditions. Such techniques include optical tweezers, atomic force microscopy, magnetocytometry, micropipetes, microplates, cell poking and particle tracking microrheology \cite{Verdier2009,Mofrad2009}.

It is also recognized that living cells combine the viscoelastic properties of known soft materials such as polymeric systems, gels, foams, suspensions, etc., but also may present properties that are associated to viscoplastic systems, since they assemble rigid bodies and fluids as main constituent parts. Furthermore, when submitted to stresses, isolated living cells can develop an active response, which contributes to the high complexity of their mechanical behaviour \cite{Pilavtepe-Celik2008}.

In order to get an expedite method to follow bacterial growth, independently of the rheological characterization of the isolated cells, we used classical rheometry, resorting to a rotational rheometer. We have monitored the growth rate of a coccoid shaped bacterial species, the human pathogen \textit{Staphylococcus aureus}. The rheological measurements allowed to characterize the viscoelastic behaviour of this system by measuring the viscosity as a function of growth time for a constant shear rate, the viscosity as a function of the shear rate at different growth times, and the elastic and the viscous modulus as a function of the angular frequency at different growth times.

\textit{S. aureus} MRSA strain COL \cite{Gill2005} was used in this study. Cultures were grown at $37^\circ$C
with aeration in tryptic soy broth (TSB) or tryptic soy agar (TSA) (Difco Laboratories, Detroit, Mich.).
Most bacteria present two distinct types of growth, the planktonic growth, a suspension of dispersed cells which occurs in liquid environment (commonly studied in laboratory, as in our work) and biofilm growth, a microbial community that adheres to a solid surface and is surrounded by a self-produced extracellular matrix (prevalent in natural environments and also in industrial and hospital settings).

Rheological measurements were performed in a Bohlin Gemini HR$^\text{nano}$ stress controlled rotational rheometer, with different geometries and applying different tests as follows:
i) steady-state shear growth curves: steel P/P geometry with  $\phi=40$ mm, gap $=2000$ $\mu$m; the viscosity was measured over time, at a constant shear rate of $10 $ s$^{-1}$, at $37^\circ$C (optimal growth temperature);
ii) steady-state shear flow curves: steel C/P geometry with $\phi=40$ mm, angle $=2^\circ$, gap $=70$ $\mu$m; the viscosity was measured as a function of the shear rate, assuring a steady-state regime in each step, imposing a minimum of 100 deformation units; measurements were performed at $20$ $^\circ$C (non optimal growth temperature);
iii) dynamic shear flow curves: steel C/P geometry with $\phi=40$ mm, angle $=2^\circ$, gap $=70$ $\mu$m; the elastic (storage) modulus, $G'$,
and the viscous (loss) modulus, $G''$, were measured as a function of the angular frequency, $\omega$, in the linear regime, imposing 10\% of strain; assays were performed at $20$ $^\circ$C. A solvent trap was used in all measurements to avoid evaporation.

\begin{figure}
\begin{center}
\includegraphics[scale=0.3]{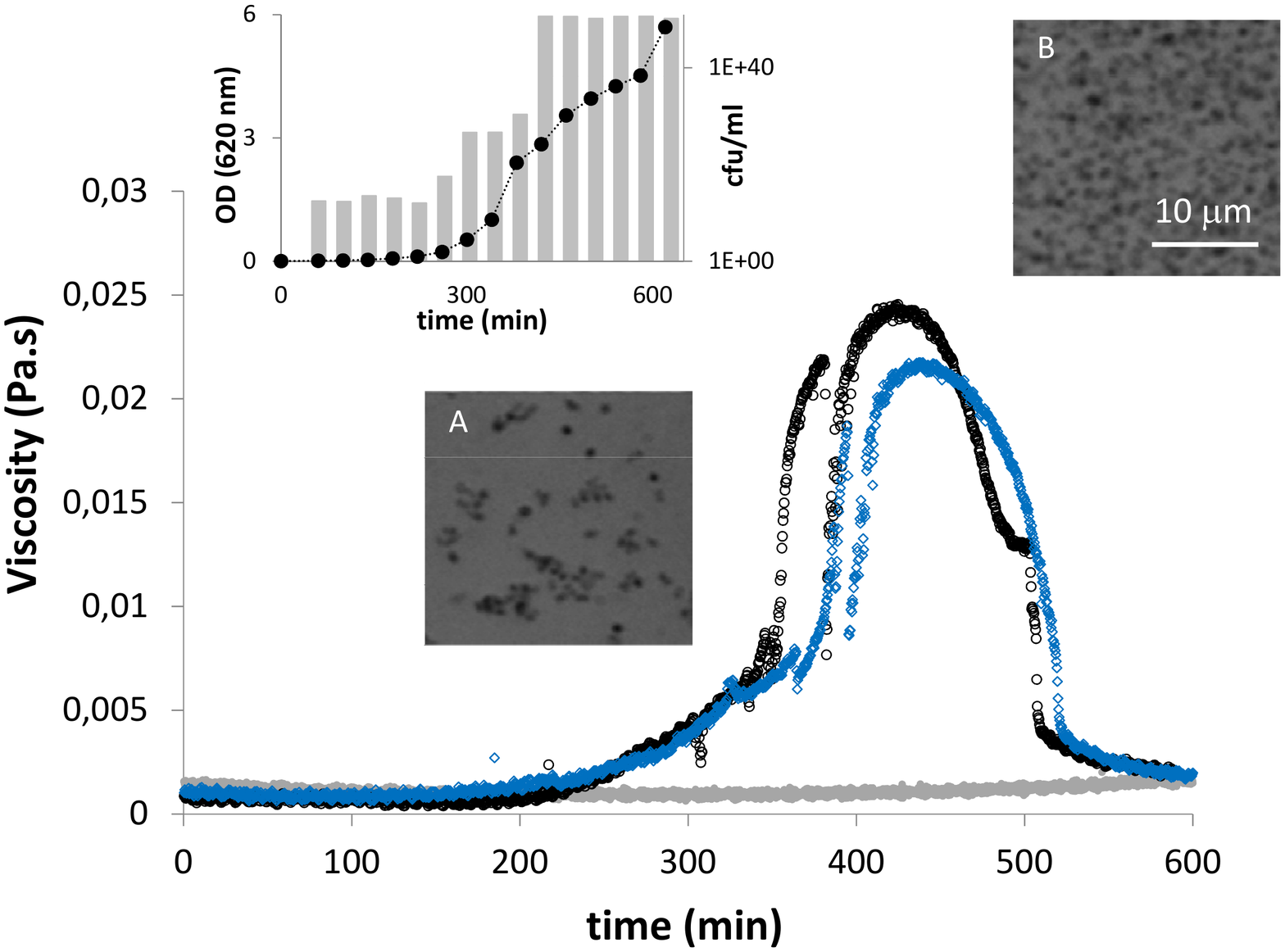}
\caption{\textit{S. aureus} culture steady-state shear growth curve, measured at a constant shear rate of
$10$ $s^{-1}$; grey-line -- culture medium; black and blue open symbols -- \textit{S. aureus} culture (representative curves); A and B -- optical microscopy images obtained during the growth curve at, respectively, 300 and 500 min;
insert: \textit{S. aureus} culture characterized by optical densities ($\text{OD}_{620\text{nm}}$) (full symbols) and population's colony forming units (cfu/ml) (bars).
All measurements were performed at 37$^\circ$C.}
\label{fig1}
\end{center}
\end{figure}

Bacterial growth was characterized in real time by rheology, during the application of a steady-state shear flow and in parallel monitored by measuring the optical density OD ($620$ nm), at discrete time intervals. Representative curves of the obtained results are included in Fig.~\ref{fig1}. The starting inocula ($OD=0.005$) was obtained from an over-night grown liquid culture. Furthermore, the population's colony forming units (cfu/ml), which provides an estimate of the viable cells, was determined over time by plating serial dilutions of the bacterial cultures on TSA, incubating for $48$ h at $37^\circ$C and counting the number of colonies.
Optical microscopy images were obtained during the growth curve with a particle characterization instrumentation (Malvern Morphologi-G3).

It is clear that there is a strong correlation between the progress of viscosity, OD and cfu/ml growth curves.
The first $300$ min of growth correspond to the lag phase, as bacteria are adapting to the new environmental growth conditions. The monitoring of OD during this time range shows a characteristic slow division rate. In accordance, the viscosity values obtained show a discrete and constant increment.
In the exponential phase of bacterial growth ($300-460$ min), the OD value increased significantly, $\sim 5\times$ and this was accompanied by a large increase in the viscosity, of $\sim 10\times$, although in a shorter time gap.
In between $450-500$ min, the $\text{OD}_{620\text{nm}}$ continued to increase, although at a slower rate, most probably due to nutrient depletion (late exponential phase). In this time window, the viscosity growth curve changed dramatically: at $\sim$ 450 min, the viscosity reached its maximum and decreased rapidly to a value close to the initial one.
Moreover, around $400$ min, a striking phenomenon occurred in the viscosity growth curve, with no observable counterpart in the optical density curve. At this moment, the viscosity suffered an instant drop and immediate recovery in a $15$ min time interval.
All assays were repeated several times and the overall profile of the viscosity growth curve was maintained, including the drops in the viscosity at similar moments (see Fig.~\ref{fig1}).

\begin{figure}
\begin{center}
\includegraphics[scale=0.3]{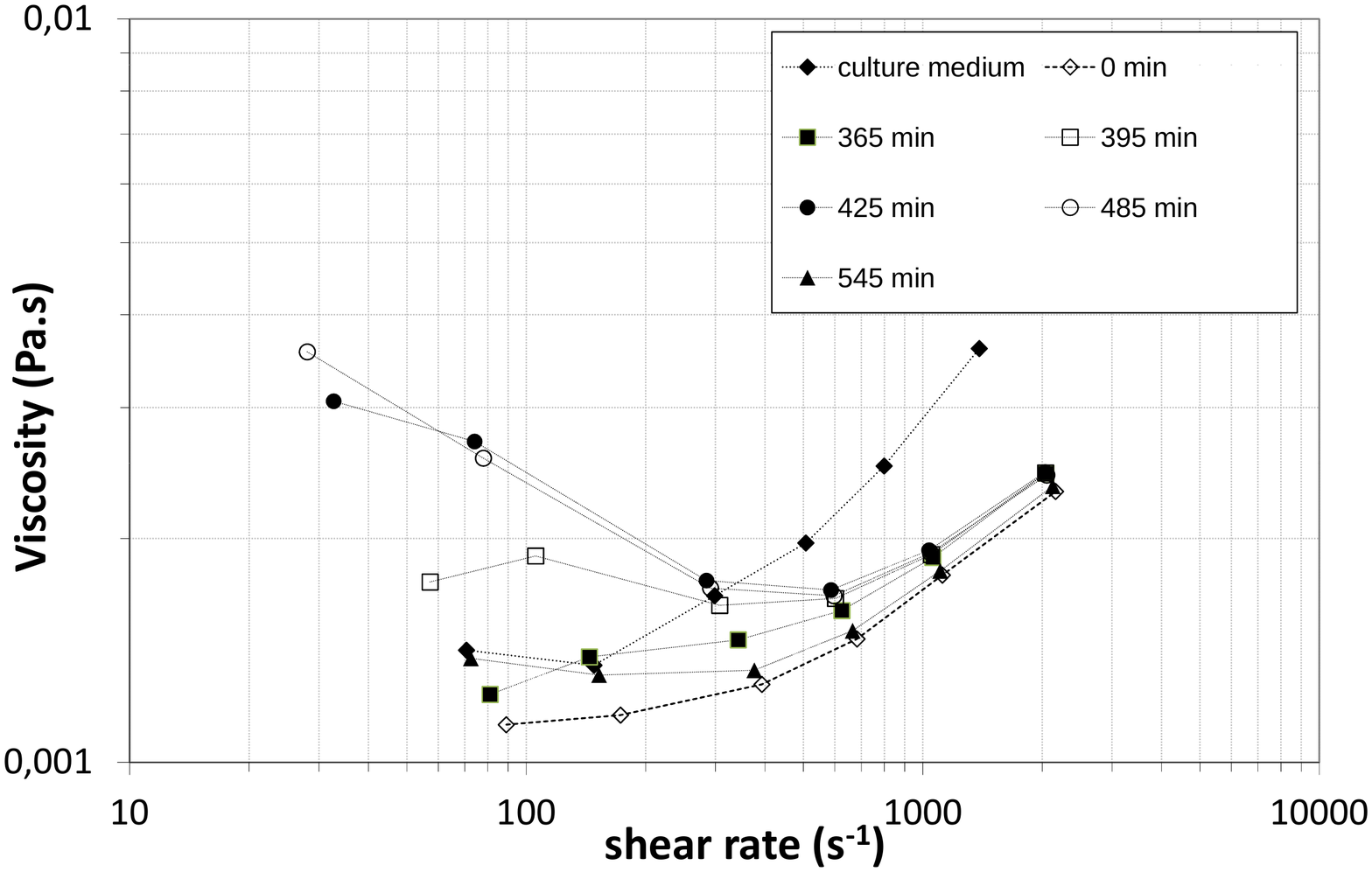}
\caption{Representative simple shear flow curves for the \textit{S. aureus} culture aliquots, measurements performed at 20$^\circ$C.}
\label{fig2}
\end{center}
\end{figure}

Simple shear flow experiments were applied to aliquots of the \textit{S. aureus} bacterial culture sampled in the beginning of the growth procedure and at the same and each time points at which the OD and cfu/ml were determined.
Representative flow curves are shown in Fig.~\ref{fig2}.

For shear rates higher than $200$ $s^{-1}$, all flow curves presented a shear thickening behaviour, comparable to the behaviour of the sterile broth medium. For lower shear rates, a shear thinning behaviour was observed, which became more evident with bacterial growth.
These results suggest several interesting features:
i) for the first aliquots, the presence of a small number of bacteria cells resulted in a decrease in the medium viscosity.
This was also observed in Fig.~\ref{fig1}, where the viscosity of the medium is slightly higher until $\sim 150$ min;
ii) for each shear rate value, the viscosity increased, reaching its maximum at $\sim 425$ min, and subsequently decreased over time; this is in agreement with the general growth curve behaviour (Fig.~\ref{fig1});
iii) in the shear rate range explored (higher than the shear rate used in Fig.~\ref{fig1}), the increase of the viscosity was never higher than $2-3\times$ (lesser than the $\sim 10\times$ viscosity increase observed in Fig.~\ref{fig1}).

\begin{figure}
\begin{center}
\includegraphics[scale=0.3]{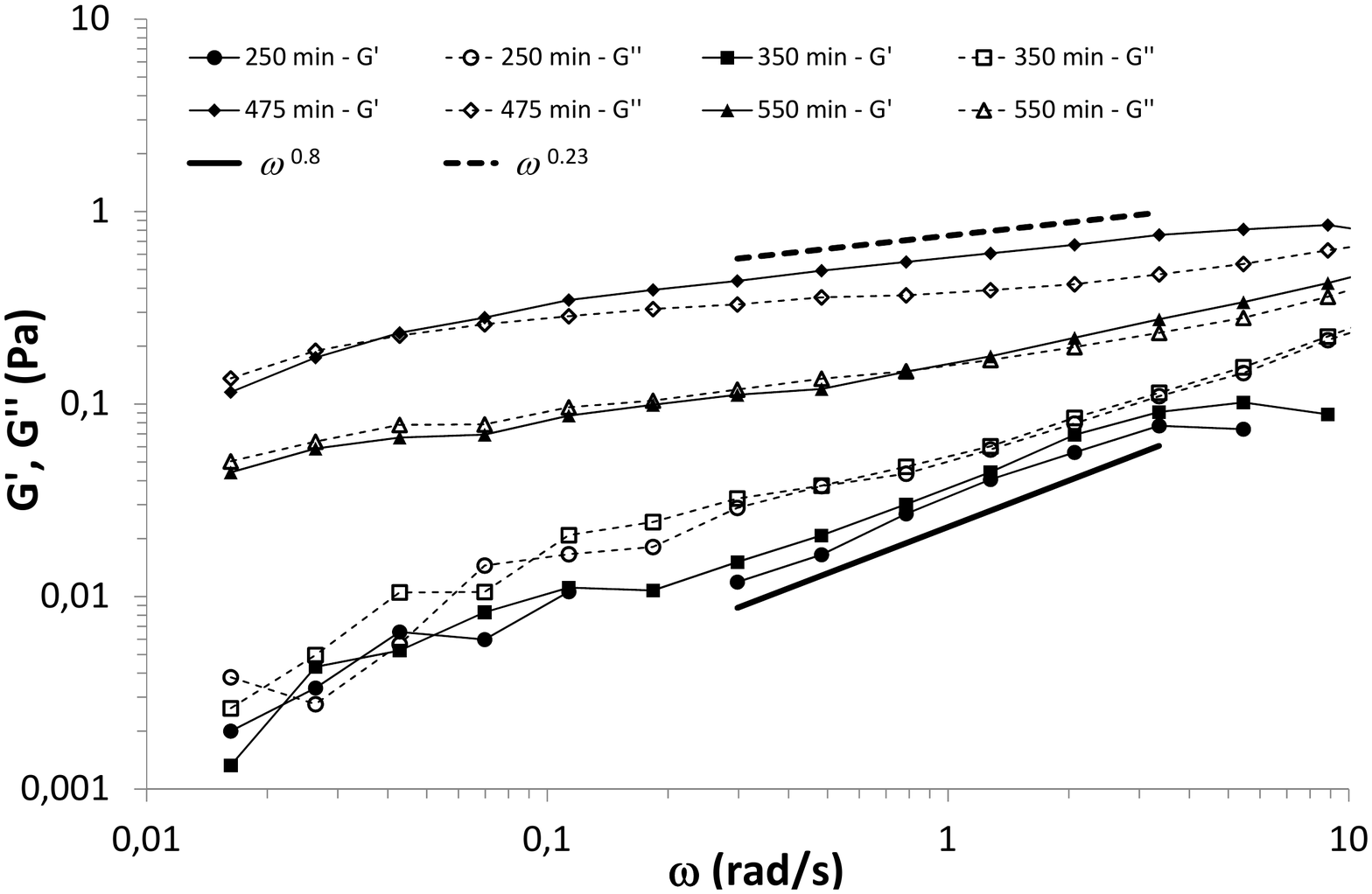}
\caption{Representative oscillatory shear flow experiments of the \textit{S. aureus} culture aliquots: elastic modulus, $G'$ and viscous modulus, $G''$, in function of the angular velocity; measurements performed at 20$^\circ$C. Solid and dashed lines correspond to the power laws of $w^{0.80}$
and $w^{0.23}$, respectively, as guides to the eye.}
\label{fig3}
\end{center}
\end{figure}

Oscillatory flow measurements were also obtained for the same culture aliquots.
The experimental values for the elastic, $G'$, and the viscous, $G''$, modulus as a function of the angular velocity, $\omega$,
are represented in Fig.~\ref{fig3}, and the following remarks can be pointed out:
i) as a general behaviour, all elastic and viscous moduli increased with $\omega$, following different but distinctive power laws;
ii) the first samples (at $250$ and $350$ min) had smaller $G'$ and $G''$, but their power law had higher exponential values. As a guide to the eye, we have represented the power law $\omega^{0.80}$, which fits well a particular part of the $250$ min - $G'$ curve;
iii) $G'$ and $G''$ increased as the bacteria grew, and reached maximal values at $475$ min. For small values of $\omega$, both $G'$ and $G''$ were $\sim 10\times$ higher than the values obtained for earlier growth stages. For large values of $\omega$, $G'$ and $G''$ increase was smaller.
In fact, as it may be seen in Fig.~\ref{fig3}, at $475$ min, $G'$ curve may be well fitted by $\omega^{0.23}$, at least in some range of $\omega$;
iv) after this stage of evolution, $G'$ and $G''$ decreased (see $550$ min curve). In turn, the associated power law exponents increased again.

To understand and explain the observed rheological results, it is important to have a good microscopic and physiological description of the bacteria behaviour, as they divide, disperse or aggregate and eventually fill the system.
At initial states, the culture may be considered as a colloid of small spheres (radius $\sim 0.5$ $\mu$m) dispersed in the liquid medium.
This type of bacteria is known to produce, specially during its exponential growth phase, adhesins, cell-surface proteins which are virulence factors and promote adhesion between bacteria (or to other cells or substrates) \cite{Tompkins1992,Voyich2005}. The regulation of the expression of virulence factors in \textit{S. aureus} is a complex and time-dependent process, being controlled by genetic regulatory systems, which respond to environmental stimuli and population density and through quorum sensing mechanisms \cite{Cheung1992,Geisinger2009,George2007,Recsei1986}.
In an infection scenario, during early stages, the expression of a vast array of adhesion factors, is up-regulated, in order to allow the bacteria to attach to the host cells. For low bacteria densities, it is common to find them in small clusters of 10-20 units,
which resemble grape-clusters.
During bacteria division, these aggregates grow and may become unstable, and small clusters separate and disperse in the medium, often as an active strategy to explore richer regions in nutrients
\footnote{
Recently, different strains of \textit{S. aureus} have been reported to assemble in larger non-attached aggregates, which are structurally very stable and metabolically active \cite{Haaber2012}.
These strains produce polysaccharide intercellular adhesin (PIA), shown to be the major structural component of the planktonic aggregates, also frequently involved in the formation of biofilms \cite{Fluckiger2005}.
Our model organism strain (COL) is not a PIA producer, and has no biofilm-forming capacity.
So, our strain should be probably dependent on other adhesive molecules such as extracellular DNA (eDNA) and cell surface proteins, 
also common components of biofilms.}.
As the density increases, the bacteria aggregates start to establish new contacts and form frequently a web or percolated structure \cite{Salek2012}.
At some point, the cell density leads to a different stage of infection, in which the expression of surface adhesion determinants is inhibited and the expression of other virulence factors, secreted to the environment, is promoted \cite{Voyich2005}.


This description explains well the main features observed in Fig.~\ref{fig1}.
As the number of bacteria present in the culture overcomes its percolation threshold, the viscosity rapidly increases
($350-400$ min). It is expected that relatively low density percolated structures, which usually block or jam the motion, unblock and allow, in small periods of time, a stress release, which corresponds to a viscosity decrease. As the density of bacteria increases, unjamming transitions become more intense, until eventually the system reaches a high enough density, for which no more stress release is expected.
The viscosity reached its maximum value at $\sim 450$ min, as the evolution of cfu/ml stabilized in its highest value.
From this moment, although the number of cfu/ml remained constant (indicating an unaltered population viability),
the viscosity dramatically decreased close to its initial value.
A change in bacteria activity, must be surely the cause of this striking behaviour.
At this stage, bacteria diminish substantially the production of adhesins.
Without being able to adhere, bacteria do not resist the flow, and viscosity decreases close to its original value.
Although the assay was continued further for several hours, the viscosity remained unaltered.
We hypothesize, that for this high cell population density, the quorum sensing mechanisms are repressing the production of adhesive factors and the cell population is no longer able to re-establish the percolated structure
\footnote{
\textit{Pseudomonas aeruginosa} also forms non-attached planktonic aggregates, although much less robust than the one described for \textit{S. aureus} \cite{Alhede2011}. The \textit{P. aeruginosa} aggregates are composed of densely packed viable cells and eDNA, which is more in line with our model system. Interestingly, upon entry in stationary phase, under starvation conditions, these aggregates become dispersed as cells begin to detach \cite{Schleheck2009}. Cell detachment would be a suitable explanation for the drop in viscosity observed in the beginning of stationary phase.}.

These ideas allow us to understand also the results of Fig.~\ref{fig2}. As the population of bacteria increases, the viscosity becomes larger, mostly for small shear rates. In fact, percolation structures are fragile and are expected to diminish as the imposed shear rate increases. So, if in the conditions of Fig.~\ref{fig1}, the viscosity increased $\sim 10\times$, 
in the conditions of Fig.~\ref{fig2} we observed only a $2-3\times$ increase for the lower shear rates, and only a minor gain for larger shear rates.

The oscillatory flow measurements shown in Fig.~\ref{fig3} cannot be explained by traditional viscoelastic models.
The possibility of having $G'$ and $G''$ with weak power-law dependencies configures a system with properties close to a glassy material where disorder and metastability are essential features. Such a system is known as a ``soft glassy material'' (SGM) \cite{Sollich1997}.
At some range of (intermediate) angular frequencies it was already observed that living cell systems also present a SGM behaviour \cite{Fabry2001,Rogers2008,Wilhelm2008}. In the SGM model, $G'\sim w^{x-1}$ and $G''\sim w^{x-1}$, where $x$ is an effective noise temperature, related to how jammed the system is. When $x=1$ the system behaves as a perfect elastic body (solidlike), when $x>1$ the system can flow and becomes disordered (fluidlike when $x=2$). The results presented in Fig.~\ref{fig3} show a fluidlike behaviour ($x=1.80$) for the first stages of growth, where we think the percolated structures start to appear. At $475$ min, the system adopted an almost solidlike behaviour ($x=1.23$).
Finally, and remarkably, the system became again more fluid.

With this study it was possible to characterize in real-time the mechanical behaviour of an actively growing \textit{S. aureus} culture by rheology. The viscoelastic properties of the \textit{S. aureus} culture suffered extensive changes along the growth time and it was possible to identify critical and reproducible behaviours that occur at specific time intervals during the growth process.
These growth-dependent behaviours, which result in the virulence characteristics of the bacteria, were consistent with the model hereby proposed of the development of percolated structures based on cell-to-cell adhesion.
It is clear that \textit{S. aureus} cellular activity, at $37$ $^\circ$C, through the development of cell organization structures leads to phenomena which are not observable by common microbiological methods but result in striking alterations of the rheological properties of the cell suspension.
Finally, in this work we evidenced the usefulness of rheological approaches to study complex population dynamics. While research efforts have focused on the study of biofilms, only scarce knowledge exists on the structure of planktonic populations and their capacity to aggregate. This is especially relevant for the study of opportunistic bacteria, as non-attached aggregates may be responsible for bacterial spreading in many clinical scenarios of persistent and chronic infections. We are presently extending our studies to differentiate the population behavior and aggregation patterns of other \textit{S. aureus} strains, with different intercellular adhesion and biofilm producing phenotypes and also of other bacterial species.

We thank J. Catita and PARALAB for the use of the particle characterization instrument Malvern Morphologi-G3.
Strain COL was a kind gift from H. de Lencastre.
We acknowledge the support from FCT (Portugal) through Grant No. PEst-OE/FIS/UI0618/2011 (CFTC), PEst-C/CTM/LA0025/2011 (CENIMAT/I3N),
PEst-OE/BIA/UI0457/2011 (CREM), and through Project PTDC/BIA/MIC/101375/2008 (awarded to RGS).

\bibliography{BacteriaGrowth}

\begin{thebibliography}{24}%
\makeatletter
\providecommand \@ifxundefined [1]{%
 \@ifx{#1\undefined}
}%
\providecommand \@ifnum [1]{%
 \ifnum #1\expandafter \@firstoftwo
 \else \expandafter \@secondoftwo
 \fi
}%
\providecommand \@ifx [1]{%
 \ifx #1\expandafter \@firstoftwo
 \else \expandafter \@secondoftwo
 \fi
}%
\providecommand \natexlab [1]{#1}%
\providecommand \enquote  [1]{``#1''}%
\providecommand \bibnamefont  [1]{#1}%
\providecommand \bibfnamefont [1]{#1}%
\providecommand \citenamefont [1]{#1}%
\providecommand \href@noop [0]{\@secondoftwo}%
\providecommand \href [0]{\begingroup \@sanitize@url \@href}%
\providecommand \@href[1]{\@@startlink{#1}\@@href}%
\providecommand \@@href[1]{\endgroup#1\@@endlink}%
\providecommand \@sanitize@url [0]{\catcode `\\12\catcode `\$12\catcode
  `\&12\catcode `\#12\catcode `\^12\catcode `\_12\catcode `\%12\relax}%
\providecommand \@@startlink[1]{}%
\providecommand \@@endlink[0]{}%
\providecommand \url  [0]{\begingroup\@sanitize@url \@url }%
\providecommand \@url [1]{\endgroup\@href {#1}{\urlprefix }}%
\providecommand \urlprefix  [0]{URL }%
\providecommand \Eprint [0]{\href }%
\providecommand \doibase [0]{http://dx.doi.org/}%
\providecommand \selectlanguage [0]{\@gobble}%
\providecommand \bibinfo  [0]{\@secondoftwo}%
\providecommand \bibfield  [0]{\@secondoftwo}%
\providecommand \translation [1]{[#1]}%
\providecommand \BibitemOpen [0]{}%
\providecommand \bibitemStop [0]{}%
\providecommand \bibitemNoStop [0]{.\EOS\space}%
\providecommand \EOS [0]{\spacefactor3000\relax}%
\providecommand \BibitemShut  [1]{\csname bibitem#1\endcsname}%
\let\auto@bib@innerbib\@empty
\bibitem [{\citenamefont {Nadell}\ \emph {et~al.}(2010)\citenamefont {Nadell},
  \citenamefont {Foster},\ and\ \citenamefont {Xavier}}]{Nadell2010}%
  \BibitemOpen
  \bibfield  {author} {\bibinfo {author} {\bibfnamefont {C.~D.}\ \bibnamefont
  {Nadell}}, \bibinfo {author} {\bibfnamefont {K.~R.}\ \bibnamefont {Foster}},
  \ and\ \bibinfo {author} {\bibfnamefont {J.~B.}\ \bibnamefont {Xavier}},\
  }\href {\doibase 10.1371/journal.pcbi.1000716} {\bibfield  {journal}
  {\bibinfo  {journal} {PLoS Comput. Biol.}\ }\textbf {\bibinfo {volume} {6}},\
  \bibinfo {pages} {e1000716} (\bibinfo {year} {2010})}\BibitemShut {NoStop}%
\bibitem [{\citenamefont {Fabry}\ \emph {et~al.}(2001)\citenamefont {Fabry},
  \citenamefont {Maksym}, \citenamefont {Butler}, \citenamefont {Glogauer},
  \citenamefont {Navajas},\ and\ \citenamefont {Fredberg}}]{Fabry2001}%
  \BibitemOpen
  \bibfield  {author} {\bibinfo {author} {\bibfnamefont {B.}~\bibnamefont
  {Fabry}}, \bibinfo {author} {\bibfnamefont {G.~N.}\ \bibnamefont {Maksym}},
  \bibinfo {author} {\bibfnamefont {J.~P.}\ \bibnamefont {Butler}}, \bibinfo
  {author} {\bibfnamefont {M.}~\bibnamefont {Glogauer}}, \bibinfo {author}
  {\bibfnamefont {D.}~\bibnamefont {Navajas}}, \ and\ \bibinfo {author}
  {\bibfnamefont {J.~J.}\ \bibnamefont {Fredberg}},\ }\href {\doibase
  10.1103/PhysRevLett.87.148102} {\bibfield  {journal} {\bibinfo  {journal}
  {Phys. Rev. Lett.}\ }\textbf {\bibinfo {volume} {87}},\ \bibinfo {pages}
  {148102} (\bibinfo {year} {2001})}\BibitemShut {NoStop}%
\bibitem [{\citenamefont {Wilhelm}(2008)}]{Wilhelm2008}%
  \BibitemOpen
  \bibfield  {author} {\bibinfo {author} {\bibfnamefont {C.}~\bibnamefont
  {Wilhelm}},\ }\href {\doibase 10.1103/PhysRevLett.101.028101} {\bibfield
  {journal} {\bibinfo  {journal} {Phys. Rev. Lett.}\ }\textbf {\bibinfo
  {volume} {101}},\ \bibinfo {pages} {028101} (\bibinfo {year}
  {2008})}\BibitemShut {NoStop}%
\bibitem [{\citenamefont {Klapper}\ \emph {et~al.}(2002)\citenamefont
  {Klapper}, \citenamefont {Rupp}, \citenamefont {Cargo}, \citenamefont
  {Purvedorj},\ and\ \citenamefont {Stoodley}}]{Klapper2002}%
  \BibitemOpen
  \bibfield  {author} {\bibinfo {author} {\bibfnamefont {I.}~\bibnamefont
  {Klapper}}, \bibinfo {author} {\bibfnamefont {C.~J.}\ \bibnamefont {Rupp}},
  \bibinfo {author} {\bibfnamefont {R.}~\bibnamefont {Cargo}}, \bibinfo
  {author} {\bibfnamefont {B.}~\bibnamefont {Purvedorj}}, \ and\ \bibinfo
  {author} {\bibfnamefont {P.}~\bibnamefont {Stoodley}},\ }\href {\doibase
  10.1002/bit.10376} {\bibfield  {journal} {\bibinfo  {journal} {Biotechnol.
  Bioeng.}\ }\textbf {\bibinfo {volume} {80}},\ \bibinfo {pages} {289}
  (\bibinfo {year} {2002})}\BibitemShut {NoStop}%
\bibitem [{\citenamefont {Rogers}\ \emph {et~al.}(2008)\citenamefont {Rogers},
  \citenamefont {van~der Walle},\ and\ \citenamefont {Waigh}}]{Rogers2008}%
  \BibitemOpen
  \bibfield  {author} {\bibinfo {author} {\bibfnamefont {S.~S.}\ \bibnamefont
  {Rogers}}, \bibinfo {author} {\bibfnamefont {C.}~\bibnamefont {van~der
  Walle}}, \ and\ \bibinfo {author} {\bibfnamefont {T.~A.}\ \bibnamefont
  {Waigh}},\ }\href {\doibase 10.1021/la802442d} {\bibfield  {journal}
  {\bibinfo  {journal} {Langmuir}\ }\textbf {\bibinfo {volume} {24}},\ \bibinfo
  {pages} {13549} (\bibinfo {year} {2008})}\BibitemShut {NoStop}%
\bibitem [{\citenamefont {Rupp}\ \emph {et~al.}(2005)\citenamefont {Rupp},
  \citenamefont {Fux},\ and\ \citenamefont {Stoodley}}]{Rupp2005}%
  \BibitemOpen
  \bibfield  {author} {\bibinfo {author} {\bibfnamefont {C.~J.}\ \bibnamefont
  {Rupp}}, \bibinfo {author} {\bibfnamefont {C.~A.}\ \bibnamefont {Fux}}, \
  and\ \bibinfo {author} {\bibfnamefont {P.}~\bibnamefont {Stoodley}},\ }\href
  {\doibase 10.1128/AEM.71.4.2175-2178.2005} {\bibfield  {journal} {\bibinfo
  {journal} {Appl. Environ. Microbiol.}\ }\textbf {\bibinfo {volume} {71}},\
  \bibinfo {pages} {2175} (\bibinfo {year} {2005})}\BibitemShut {NoStop}%
\bibitem [{\citenamefont {Verdier}\ \emph {et~al.}(2009)\citenamefont
  {Verdier}, \citenamefont {Etienne}, \citenamefont {Duperray},\ and\
  \citenamefont {Preziosi}}]{Verdier2009}%
  \BibitemOpen
  \bibfield  {author} {\bibinfo {author} {\bibfnamefont {C.}~\bibnamefont
  {Verdier}}, \bibinfo {author} {\bibfnamefont {J.}~\bibnamefont {Etienne}},
  \bibinfo {author} {\bibfnamefont {A.}~\bibnamefont {Duperray}}, \ and\
  \bibinfo {author} {\bibfnamefont {L.}~\bibnamefont {Preziosi}},\ }\href
  {\doibase 10.1016/j.crhy.2009.10.003} {\bibfield  {journal} {\bibinfo
  {journal} {C. R. Phys.}\ }\textbf {\bibinfo {volume} {10}},\ \bibinfo {pages}
  {790} (\bibinfo {year} {2009})}\BibitemShut {NoStop}%
\bibitem [{\citenamefont {Mofrad}(2009)}]{Mofrad2009}%
  \BibitemOpen
  \bibfield  {author} {\bibinfo {author} {\bibfnamefont {M.~R.~K.}\
  \bibnamefont {Mofrad}},\ }\href {\doibase
  10.1146/annurev.fluid.010908.165236} {\bibfield  {journal} {\bibinfo
  {journal} {Annual Review of Fluid Mechanics}\ }\textbf {\bibinfo {volume}
  {41}},\ \bibinfo {pages} {433} (\bibinfo {year} {2009})}\BibitemShut
  {NoStop}%
\bibitem [{\citenamefont {Pilavtepe-Celik}\ \emph {et~al.}(2008)\citenamefont
  {Pilavtepe-Celik}, \citenamefont {Balaban}, \citenamefont {Alpas},\ and\
  \citenamefont {Yousef}}]{Pilavtepe-Celik2008}%
  \BibitemOpen
  \bibfield  {author} {\bibinfo {author} {\bibfnamefont {M.}~\bibnamefont
  {Pilavtepe-Celik}}, \bibinfo {author} {\bibfnamefont {M.~O.}\ \bibnamefont
  {Balaban}}, \bibinfo {author} {\bibfnamefont {H.}~\bibnamefont {Alpas}}, \
  and\ \bibinfo {author} {\bibfnamefont {A.~E.}\ \bibnamefont {Yousef}},\
  }\href {\doibase 10.1111/j.1750-3841.2008.00947.x} {\bibfield  {journal}
  {\bibinfo  {journal} {J. Food Sci.}\ }\textbf {\bibinfo {volume} {73}},\
  \bibinfo {pages} {M423} (\bibinfo {year} {2008})}\BibitemShut {NoStop}%
\bibitem [{\citenamefont {Gill}\ \emph {et~al.}(2005)\citenamefont {Gill},
  \citenamefont {Fouts}, \citenamefont {Archer}, \citenamefont {Mongodin},
  \citenamefont {DeBoy}, \citenamefont {Ravel}, \citenamefont {Paulsen},
  \citenamefont {Kolonay}, \citenamefont {Brinkac}, \citenamefont {Beanan},
  \citenamefont {Dodson}, \citenamefont {Daugherty}, \citenamefont {Madupu},
  \citenamefont {Angiuoli}, \citenamefont {Durkin}, \citenamefont {Haft},
  \citenamefont {Vamathevan}, \citenamefont {Khouri}, \citenamefont
  {Utterback}, \citenamefont {Lee}, \citenamefont {Dimitrov}, \citenamefont
  {Jiang}, \citenamefont {Qin}, \citenamefont {Weidman}, \citenamefont {Tran},
  \citenamefont {Kang}, \citenamefont {Hance}, \citenamefont {Nelson},\ and\
  \citenamefont {Fraser}}]{Gill2005}%
  \BibitemOpen
  \bibfield  {author} {\bibinfo {author} {\bibfnamefont {S.~R.}\ \bibnamefont
  {Gill}}, \bibinfo {author} {\bibfnamefont {D.~E.}\ \bibnamefont {Fouts}},
  \bibinfo {author} {\bibfnamefont {G.~L.}\ \bibnamefont {Archer}}, \bibinfo
  {author} {\bibfnamefont {E.~F.}\ \bibnamefont {Mongodin}}, \bibinfo {author}
  {\bibfnamefont {R.~T.}\ \bibnamefont {DeBoy}}, \bibinfo {author}
  {\bibfnamefont {J.}~\bibnamefont {Ravel}}, \bibinfo {author} {\bibfnamefont
  {I.~T.}\ \bibnamefont {Paulsen}}, \bibinfo {author} {\bibfnamefont {J.~F.}\
  \bibnamefont {Kolonay}}, \bibinfo {author} {\bibfnamefont {L.}~\bibnamefont
  {Brinkac}}, \bibinfo {author} {\bibfnamefont {M.}~\bibnamefont {Beanan}},
  \bibinfo {author} {\bibfnamefont {R.~J.}\ \bibnamefont {Dodson}}, \bibinfo
  {author} {\bibfnamefont {S.~C.}\ \bibnamefont {Daugherty}}, \bibinfo {author}
  {\bibfnamefont {R.}~\bibnamefont {Madupu}}, \bibinfo {author} {\bibfnamefont
  {S.~V.}\ \bibnamefont {Angiuoli}}, \bibinfo {author} {\bibfnamefont {A.~S.}\
  \bibnamefont {Durkin}}, \bibinfo {author} {\bibfnamefont {D.~H.}\
  \bibnamefont {Haft}}, \bibinfo {author} {\bibfnamefont {J.}~\bibnamefont
  {Vamathevan}}, \bibinfo {author} {\bibfnamefont {H.}~\bibnamefont {Khouri}},
  \bibinfo {author} {\bibfnamefont {T.}~\bibnamefont {Utterback}}, \bibinfo
  {author} {\bibfnamefont {C.}~\bibnamefont {Lee}}, \bibinfo {author}
  {\bibfnamefont {G.}~\bibnamefont {Dimitrov}}, \bibinfo {author}
  {\bibfnamefont {L.~X.}\ \bibnamefont {Jiang}}, \bibinfo {author}
  {\bibfnamefont {H.~Y.}\ \bibnamefont {Qin}}, \bibinfo {author} {\bibfnamefont
  {J.}~\bibnamefont {Weidman}}, \bibinfo {author} {\bibfnamefont
  {K.}~\bibnamefont {Tran}}, \bibinfo {author} {\bibfnamefont {K.}~\bibnamefont
  {Kang}}, \bibinfo {author} {\bibfnamefont {I.~R.}\ \bibnamefont {Hance}},
  \bibinfo {author} {\bibfnamefont {K.~E.}\ \bibnamefont {Nelson}}, \ and\
  \bibinfo {author} {\bibfnamefont {C.~M.}\ \bibnamefont {Fraser}},\ }\href
  {\doibase 10.1128/JB.187.7.2426-2438.2005} {\bibfield  {journal} {\bibinfo
  {journal} {J. Bacteriol.}\ }\textbf {\bibinfo {volume} {187}},\ \bibinfo
  {pages} {2426} (\bibinfo {year} {2005})}\BibitemShut {NoStop}%
\bibitem [{\citenamefont {Tompkins}\ \emph {et~al.}(1992)\citenamefont
  {Tompkins}, \citenamefont {Blackwell}, \citenamefont {Hatcher}, \citenamefont
  {Elliott}, \citenamefont {Ohagansotsky},\ and\ \citenamefont
  {Lowy}}]{Tompkins1992}%
  \BibitemOpen
  \bibfield  {author} {\bibinfo {author} {\bibfnamefont {D.~C.}\ \bibnamefont
  {Tompkins}}, \bibinfo {author} {\bibfnamefont {L.~J.}\ \bibnamefont
  {Blackwell}}, \bibinfo {author} {\bibfnamefont {V.~B.}\ \bibnamefont
  {Hatcher}}, \bibinfo {author} {\bibfnamefont {D.~A.}\ \bibnamefont
  {Elliott}}, \bibinfo {author} {\bibfnamefont {C.}~\bibnamefont
  {Ohagansotsky}}, \ and\ \bibinfo {author} {\bibfnamefont {F.~D.}\
  \bibnamefont {Lowy}},\ }\href@noop {} {\bibfield  {journal} {\bibinfo
  {journal} {Infect. Immun.}\ }\textbf {\bibinfo {volume} {60}},\ \bibinfo
  {pages} {965} (\bibinfo {year} {1992})}\BibitemShut {NoStop}%
\bibitem [{\citenamefont {Voyich}\ \emph {et~al.}(2005)\citenamefont {Voyich},
  \citenamefont {Braughton}, \citenamefont {Sturdevant}, \citenamefont
  {Whitney}, \citenamefont {Said-Salim}, \citenamefont {Porcella},
  \citenamefont {Long}, \citenamefont {Dorward}, \citenamefont {Gardner},
  \citenamefont {Kreiswirth}, \citenamefont {Musser},\ and\ \citenamefont
  {DeLeo}}]{Voyich2005}%
  \BibitemOpen
  \bibfield  {author} {\bibinfo {author} {\bibfnamefont {J.~A.}\ \bibnamefont
  {Voyich}}, \bibinfo {author} {\bibfnamefont {K.~R.}\ \bibnamefont
  {Braughton}}, \bibinfo {author} {\bibfnamefont {D.~E.}\ \bibnamefont
  {Sturdevant}}, \bibinfo {author} {\bibfnamefont {A.~R.}\ \bibnamefont
  {Whitney}}, \bibinfo {author} {\bibfnamefont {B.}~\bibnamefont {Said-Salim}},
  \bibinfo {author} {\bibfnamefont {S.~F.}\ \bibnamefont {Porcella}}, \bibinfo
  {author} {\bibfnamefont {R.~D.}\ \bibnamefont {Long}}, \bibinfo {author}
  {\bibfnamefont {D.~W.}\ \bibnamefont {Dorward}}, \bibinfo {author}
  {\bibfnamefont {D.~J.}\ \bibnamefont {Gardner}}, \bibinfo {author}
  {\bibfnamefont {B.~N.}\ \bibnamefont {Kreiswirth}}, \bibinfo {author}
  {\bibfnamefont {J.~M.}\ \bibnamefont {Musser}}, \ and\ \bibinfo {author}
  {\bibfnamefont {F.~R.}\ \bibnamefont {DeLeo}},\ }\href@noop {} {\bibfield
  {journal} {\bibinfo  {journal} {J. Immunol.}\ }\textbf {\bibinfo {volume}
  {175}},\ \bibinfo {pages} {3907} (\bibinfo {year} {2005})}\BibitemShut
  {NoStop}%
\bibitem [{\citenamefont {Cheung}\ \emph {et~al.}(1992)\citenamefont {Cheung},
  \citenamefont {Koomey}, \citenamefont {Butler}, \citenamefont {Projan},\ and\
  \citenamefont {Fischetti}}]{Cheung1992}%
  \BibitemOpen
  \bibfield  {author} {\bibinfo {author} {\bibfnamefont {A.~L.}\ \bibnamefont
  {Cheung}}, \bibinfo {author} {\bibfnamefont {J.~M.}\ \bibnamefont {Koomey}},
  \bibinfo {author} {\bibfnamefont {C.~A.}\ \bibnamefont {Butler}}, \bibinfo
  {author} {\bibfnamefont {S.~J.}\ \bibnamefont {Projan}}, \ and\ \bibinfo
  {author} {\bibfnamefont {V.~A.}\ \bibnamefont {Fischetti}},\ }\href {\doibase
  10.1073/pnas.89.14.6462} {\bibfield  {journal} {\bibinfo  {journal} {Proc.
  Natl. Acad. Sci. U. S. A.}\ }\textbf {\bibinfo {volume} {89}},\ \bibinfo
  {pages} {6462} (\bibinfo {year} {1992})}\BibitemShut {NoStop}%
\bibitem [{\citenamefont {Geisinger}\ \emph {et~al.}(2009)\citenamefont
  {Geisinger}, \citenamefont {Muir},\ and\ \citenamefont
  {Novick}}]{Geisinger2009}%
  \BibitemOpen
  \bibfield  {author} {\bibinfo {author} {\bibfnamefont {E.}~\bibnamefont
  {Geisinger}}, \bibinfo {author} {\bibfnamefont {T.~W.}\ \bibnamefont {Muir}},
  \ and\ \bibinfo {author} {\bibfnamefont {R.~P.}\ \bibnamefont {Novick}},\
  }\href {\doibase 10.1073/pnas.0807760106} {\bibfield  {journal} {\bibinfo
  {journal} {Proc. Natl. Acad. Sci. U. S. A.}\ }\textbf {\bibinfo {volume}
  {106}},\ \bibinfo {pages} {1216} (\bibinfo {year} {2009})}\BibitemShut
  {NoStop}%
\bibitem [{\citenamefont {George}\ and\ \citenamefont
  {Muir}(2007)}]{George2007}%
  \BibitemOpen
  \bibfield  {author} {\bibinfo {author} {\bibfnamefont {E.~A.}\ \bibnamefont
  {George}}\ and\ \bibinfo {author} {\bibfnamefont {T.~W.}\ \bibnamefont
  {Muir}},\ }\href {\doibase 10.1002/cbic.200700023} {\bibfield  {journal}
  {\bibinfo  {journal} {ChemBioChem}\ }\textbf {\bibinfo {volume} {8}},\
  \bibinfo {pages} {847} (\bibinfo {year} {2007})}\BibitemShut {NoStop}%
\bibitem [{\citenamefont {Recsei}\ \emph {et~al.}(1986)\citenamefont {Recsei},
  \citenamefont {Kreiswirth}, \citenamefont {Oreilly}, \citenamefont
  {Schlievert}, \citenamefont {Gruss},\ and\ \citenamefont
  {Novick}}]{Recsei1986}%
  \BibitemOpen
  \bibfield  {author} {\bibinfo {author} {\bibfnamefont {P.}~\bibnamefont
  {Recsei}}, \bibinfo {author} {\bibfnamefont {B.}~\bibnamefont {Kreiswirth}},
  \bibinfo {author} {\bibfnamefont {M.}~\bibnamefont {Oreilly}}, \bibinfo
  {author} {\bibfnamefont {P.}~\bibnamefont {Schlievert}}, \bibinfo {author}
  {\bibfnamefont {A.}~\bibnamefont {Gruss}}, \ and\ \bibinfo {author}
  {\bibfnamefont {R.~P.}\ \bibnamefont {Novick}},\ }\href {\doibase
  10.1007/BF00330517} {\bibfield  {journal} {\bibinfo  {journal} {Molecular \&
  General Genetics}\ }\textbf {\bibinfo {volume} {202}},\ \bibinfo {pages} {58}
  (\bibinfo {year} {1986})}\BibitemShut {NoStop}%
\bibitem [{Note1()}]{Note1}%
  \BibitemOpen
  \bibinfo {note} {Recently, different strains of \protect \textit {S. aureus}
  have been reported to assemble in larger non-attached aggregates, which are
  structurally very stable and metabolically active \cite {Haaber2012}. These
  strains produce polysaccharide intercellular adhesin (PIA), shown to be the
  major structural component of the planktonic aggregates, also frequently
  involved in the formation of biofilms \cite {Fluckiger2005}. Our model
  organism strain (COL) is not a PIA producer, and has no biofilm-forming
  capacity. So, our strain should be probably dependent on other adhesive
  molecules such as extracellular DNA (eDNA) and cell surface proteins, also
  common components of biofilms.}\BibitemShut {Stop}%
\bibitem [{\citenamefont {Salek}\ \emph {et~al.}(2012)\citenamefont {Salek},
  \citenamefont {Sattari},\ and\ \citenamefont {Martinuzzi}}]{Salek2012}%
  \BibitemOpen
  \bibfield  {author} {\bibinfo {author} {\bibfnamefont {M.~M.}\ \bibnamefont
  {Salek}}, \bibinfo {author} {\bibfnamefont {P.}~\bibnamefont {Sattari}}, \
  and\ \bibinfo {author} {\bibfnamefont {R.~J.}\ \bibnamefont {Martinuzzi}},\
  }\href {\doibase 10.1007/s10439-011-0444-9} {\bibfield  {journal} {\bibinfo
  {journal} {Annals of Biomedical Engineering}\ }\textbf {\bibinfo {volume}
  {40}},\ \bibinfo {pages} {707} (\bibinfo {year} {2012})}\BibitemShut
  {NoStop}%
\bibitem [{Note2()}]{Note2}%
  \BibitemOpen
  \bibinfo {note} {\protect \textit {Pseudomonas aeruginosa} also forms
  non-attached planktonic aggregates, although much less robust than the one
  described for \protect \textit {S. aureus} \cite {Alhede2011}. The \protect
  \textit {P. aeruginosa} aggregates are composed of densely packed viable
  cells and eDNA, which is more in line with our model system. Interestingly,
  upon entry in stationary phase, under starvation conditions, these aggregates
  become dispersed as cells begin to detach \cite {Schleheck2009}. Cell
  detachment would be a suitable explanation for the drop in viscosity observed
  in the beginning of stationary phase.}\BibitemShut {Stop}%
\bibitem [{\citenamefont {Sollich}\ \emph {et~al.}(1997)\citenamefont
  {Sollich}, \citenamefont {Lequeux}, \citenamefont {Hebraud},\ and\
  \citenamefont {Cates}}]{Sollich1997}%
  \BibitemOpen
  \bibfield  {author} {\bibinfo {author} {\bibfnamefont {P.}~\bibnamefont
  {Sollich}}, \bibinfo {author} {\bibfnamefont {F.}~\bibnamefont {Lequeux}},
  \bibinfo {author} {\bibfnamefont {P.}~\bibnamefont {Hebraud}}, \ and\
  \bibinfo {author} {\bibfnamefont {M.~E.}\ \bibnamefont {Cates}},\ }\href
  {\doibase 10.1103/PhysRevLett.78.2020} {\bibfield  {journal} {\bibinfo
  {journal} {Phys. Rev. Lett.}\ }\textbf {\bibinfo {volume} {78}},\ \bibinfo
  {pages} {2020} (\bibinfo {year} {1997})}\BibitemShut {NoStop}%
\bibitem [{\citenamefont {Haaber}\ \emph {et~al.}(2012)\citenamefont {Haaber},
  \citenamefont {Cohn}, \citenamefont {Frees}, \citenamefont {Andersen},\ and\
  \citenamefont {Ingmer}}]{Haaber2012}%
  \BibitemOpen
  \bibfield  {author} {\bibinfo {author} {\bibfnamefont {J.}~\bibnamefont
  {Haaber}}, \bibinfo {author} {\bibfnamefont {M.~T.}\ \bibnamefont {Cohn}},
  \bibinfo {author} {\bibfnamefont {D.}~\bibnamefont {Frees}}, \bibinfo
  {author} {\bibfnamefont {T.~J.}\ \bibnamefont {Andersen}}, \ and\ \bibinfo
  {author} {\bibfnamefont {H.}~\bibnamefont {Ingmer}},\ }\href {\doibase
  10.1371/journal.pone.0041075} {\bibfield  {journal} {\bibinfo  {journal}
  {Plos One}\ }\textbf {\bibinfo {volume} {7}},\ \bibinfo {pages} {e41075}
  (\bibinfo {year} {2012})}\BibitemShut {NoStop}%
\bibitem [{\citenamefont {Fluckiger}\ \emph {et~al.}(2005)\citenamefont
  {Fluckiger}, \citenamefont {Ulrich}, \citenamefont {Steinhuber},
  \citenamefont {Doring}, \citenamefont {Mack}, \citenamefont {Landmann},
  \citenamefont {Goerke},\ and\ \citenamefont {Wolz}}]{Fluckiger2005}%
  \BibitemOpen
  \bibfield  {author} {\bibinfo {author} {\bibfnamefont {U.}~\bibnamefont
  {Fluckiger}}, \bibinfo {author} {\bibfnamefont {M.}~\bibnamefont {Ulrich}},
  \bibinfo {author} {\bibfnamefont {A.}~\bibnamefont {Steinhuber}}, \bibinfo
  {author} {\bibfnamefont {G.}~\bibnamefont {Doring}}, \bibinfo {author}
  {\bibfnamefont {D.}~\bibnamefont {Mack}}, \bibinfo {author} {\bibfnamefont
  {R.}~\bibnamefont {Landmann}}, \bibinfo {author} {\bibfnamefont
  {C.}~\bibnamefont {Goerke}}, \ and\ \bibinfo {author} {\bibfnamefont
  {C.}~\bibnamefont {Wolz}},\ }\href {\doibase 10.1128/IAI.73.3.1811-1819.2005}
  {\bibfield  {journal} {\bibinfo  {journal} {Infect. Immun.}\ }\textbf
  {\bibinfo {volume} {73}},\ \bibinfo {pages} {1811} (\bibinfo {year}
  {2005})}\BibitemShut {NoStop}%
\bibitem [{\citenamefont {Alhede}\ \emph {et~al.}(2011)\citenamefont {Alhede},
  \citenamefont {Kragh}, \citenamefont {Qvortrup}, \citenamefont
  {Allesen-Holm}, \citenamefont {van Gennip}, \citenamefont {Christensen},
  \citenamefont {Jensen}, \citenamefont {Nielsen}, \citenamefont {Parsek},
  \citenamefont {Wozniak}, \citenamefont {Molin}, \citenamefont
  {Tolker-Nielsen}, \citenamefont {Hoiby}, \citenamefont {Givskov},\ and\
  \citenamefont {Bjarnsholt}}]{Alhede2011}%
  \BibitemOpen
  \bibfield  {author} {\bibinfo {author} {\bibfnamefont {M.}~\bibnamefont
  {Alhede}}, \bibinfo {author} {\bibfnamefont {K.~N.}\ \bibnamefont {Kragh}},
  \bibinfo {author} {\bibfnamefont {K.}~\bibnamefont {Qvortrup}}, \bibinfo
  {author} {\bibfnamefont {M.}~\bibnamefont {Allesen-Holm}}, \bibinfo {author}
  {\bibfnamefont {M.}~\bibnamefont {van Gennip}}, \bibinfo {author}
  {\bibfnamefont {L.~D.}\ \bibnamefont {Christensen}}, \bibinfo {author}
  {\bibfnamefont {P.~O.}\ \bibnamefont {Jensen}}, \bibinfo {author}
  {\bibfnamefont {A.~K.}\ \bibnamefont {Nielsen}}, \bibinfo {author}
  {\bibfnamefont {M.}~\bibnamefont {Parsek}}, \bibinfo {author} {\bibfnamefont
  {D.}~\bibnamefont {Wozniak}}, \bibinfo {author} {\bibfnamefont
  {S.}~\bibnamefont {Molin}}, \bibinfo {author} {\bibfnamefont
  {T.}~\bibnamefont {Tolker-Nielsen}}, \bibinfo {author} {\bibfnamefont
  {N.}~\bibnamefont {Hoiby}}, \bibinfo {author} {\bibfnamefont
  {M.}~\bibnamefont {Givskov}}, \ and\ \bibinfo {author} {\bibfnamefont
  {T.}~\bibnamefont {Bjarnsholt}},\ }\href {\doibase
  10.1371/journal.pone.0027943} {\bibfield  {journal} {\bibinfo  {journal}
  {Plos One}\ }\textbf {\bibinfo {volume} {6}},\ \bibinfo {pages} {e27943}
  (\bibinfo {year} {2011})}\BibitemShut {NoStop}%
\bibitem [{\citenamefont {Schleheck}\ \emph {et~al.}(2009)\citenamefont
  {Schleheck}, \citenamefont {Barraud}, \citenamefont {Klebensberger},
  \citenamefont {Webb}, \citenamefont {McDougald}, \citenamefont {Rice},\ and\
  \citenamefont {Kjelleberg}}]{Schleheck2009}%
  \BibitemOpen
  \bibfield  {author} {\bibinfo {author} {\bibfnamefont {D.}~\bibnamefont
  {Schleheck}}, \bibinfo {author} {\bibfnamefont {N.}~\bibnamefont {Barraud}},
  \bibinfo {author} {\bibfnamefont {J.}~\bibnamefont {Klebensberger}}, \bibinfo
  {author} {\bibfnamefont {J.~S.}\ \bibnamefont {Webb}}, \bibinfo {author}
  {\bibfnamefont {D.}~\bibnamefont {McDougald}}, \bibinfo {author}
  {\bibfnamefont {S.~A.}\ \bibnamefont {Rice}}, \ and\ \bibinfo {author}
  {\bibfnamefont {S.}~\bibnamefont {Kjelleberg}},\ }\href {\doibase
  10.1371/journal.pone.0005513} {\bibfield  {journal} {\bibinfo  {journal}
  {Plos One}\ }\textbf {\bibinfo {volume} {4}},\ \bibinfo {pages} {e5513}
  (\bibinfo {year} {2009})}\BibitemShut {NoStop}%
\end{thebibliography}%

\end{document}